\documentstyle[epsf]{l-aa-ps}
\def\rerg{\rm erg}
\def\rs{\rm s}
\def\rs1{\rm s^{-1}}
\def\fxg{f_{X/\gamma} }

\def\rcm{\rm cm}
\def\rcm2{\rm cm^{-2}}
\def\deg{\rm ^{\circ}}
\def\flux{\rerg\ \rcm2\ \rs1}
\def\rkeV{\rm keV}

\def\etal{et al.}

\begin{document}

\thesaurus{011
           (13.07.1;   
            13.25.1;)} 

\title{Evidence for a late-time outburst of the X$-$ray afterglow of GB970508
from BeppoSAX}

\author{ L. Piro \inst{1},
L. Amati \inst{1,10}, 
L. A. Antonelli \inst{2},
R. C. Butler \inst{3},
E. Costa \inst{1},
G. Cusumano \inst{4},
M. Feroci \inst{1},
F. Frontera \inst{5,8},
J. Heise \inst{6},
J. J. M. in 't Zand \inst{6},
S. Molendi \inst{7},
J. Muller \inst{2,6},
L. Nicastro \inst{8},
M. Orlandini \inst{8},
A. Owens \inst{9},
A.N. Parmar \inst{9},
P. Soffitta \inst{1},
M. Tavani \inst{7,11}
}

\institute{
{Istituto Astrofisica Spaziale, C.N.R., Via Fosso del Cavaliere, 00131 Roma, Italy}
\and
{BeppoSAX Science Data Center,  Via Corcolle 19, 00131 
Roma, Italy }
\and
{Agenzia Spaziale Italiana, V.le R. Margherita 120, Roma, Italy}
\and
{Istituto Fisica Cosmica e Applicazioni Calcolo Informatico, Palermo}
\and
{Dip. Fisica, Universit\`a Ferrara, Via Paradiso 12, Ferrara, Italy} 
\and
{Space Research Organization  Netherlands, 
Sorbonnelaan 2, 3584 CA Utrecht, The Netherlands} \and
{Istituto Fisica Cosmica e Tecnologie Relative, C.N.R., Milano}
\and
{Istituto TeSRE, C.N.R., Via Gobetti 101, 40129 Bologna, 
Italy}
\and
{Space Science Department of ESA, ESTEC PO 299 2200AG Noordwijk, The Netherlands}
\and
{Istituto Astronomico, Universit\`a degli Studi ``La Sapienza'', via Lancisi 29, Rome, Italy}
\and
{Columbia Astrophysics Laboratory, Columbia University, New York, 10027, USA}
}

\offprints{L. Piro:piro@alpha1.ias.rm.cnr.it}
\date{ }
\maketitle
\markboth{Piro et al.: Evidence for a late-time outburst of the X$-$ray
 afterglow of GB970508 from BeppoSAX}{}
\begin{abstract}
The $\gamma-$ray burst 
GB970508  was observed simultaneously  by the Gamma Ray Burst Monitor 
(GRBM) and one of the X$-$ray Wide Field Cameras (WFC) aboard BeppoSAX.
The latter  provided a position within $1.9'$ radius.
A series of follow-up observations with the Narrow Field Intruments (NFI) was 
then performed in a period from $\sim 6$ hours to 6 days after the 
main event. 
A previously unknown source, which we associate
with the afterglow of the GRB,
was discovered in the error box. 
We find that, after the initial burst, 
X$-$ray emission is still present and decays as
$\sim t^{-1.1}$  
up to $\sim 6\times 10^4$ s.
This is followed by 
a burst of 
activity with a duration $\sim 10^5$ s. 
The energy produced in this event 
is a substantial fraction of the total energy of the GRB, which
means that 
the  afterglow is not  
a remnant of the initial burst (the GRB) that
fades away smoothly. 
Our results  support the idea that the processes
generating the GRB and its afterglow are the same.
\end{abstract}
\section{ Introduction }
The BeppoSAX\footnote{BeppoSAX is a program of the Italian Space Agency (ASI)
 with participation of the 
Netherlands Agency for Aerospace Program (NIVR)}
(Piro, Scarsi \& Butler 1995, Boella et al. 1997a)
 observations of GB970228
(\cite{c97nat}) opened a new era in the study of GRB's
with the first discovery of an X$-$ray afterglow of GRB, 
followed by
at least three well established similar detections
 in GB970402 (Piro et al. 1997a), GB970508 (Piro et al. 1997b) and then GB970828 by XTE/ASCA (Murakami et al. 1997).
 Other possible X$-$ray afterglow candidates include GB970111 (Feroci et al. 1997), GB970616 (Marshall et al. 1997), GB970815 (Greiner et al. 1997).
 In this paper, 
we present the BeppoSAX observations of
GB970508.
Its X$-$ray evolution is tracked from
1  to $10^6$ s, i.e. from the initial burst to the
afterglow.
We compare it with that observed in other GRB's
and discuss some of the implications
on current models.

\section{The observations}

The GRBM (\cite{grbm})
 was triggered on May 8 1997  at 21:41:50  U.T. by a GRB,
 also observed by BATSE (Kouveliotou et al. 1997) and Ulysses.  
The event was simultaneously detected in one of the WFC
(\cite{wfc}). A first preliminary ($\sim 10'$)
position was derived (\cite{Costa97}) and used to 
program a follow-up observation
with the NFI. Simultaneously this position (followed then by a refined
 $3'$ one
(\cite{H97}) and the $50"$  derived from NFI (\cite{Piro97})
were distributed  to a network of observatories
for follow-up observations in all wavelengths. This led to the identification of
an optical transient just 4 hours after the burst (\cite{Bond97}) and eventually to the spectroscopic observation that set the distance of the optical transient at 
$z>0.83$ (\cite{MTZ97}).

The field was acquired by the NFI $\sim 6$ hours after the GRB.
A previously unknown X$-$ray source, 1SAX J0653.8+7916 was detected
in this observation (hereafter TOO1) 
by the MECS (units 2 and 3) (\cite{mecs}) 
with F(2$-$10 keV)=$(0.7\pm0.07)\times 10^{-12} \flux$
and the 
LECS (\cite{lecs}) with F(0.1$-$2 keV)=$ (1.2\pm0.4)\times
10^{-12} \flux$ at celestial coordinates (J2000)
R.A.=6h53m46s.7, Decl.=$+79\deg16'02"$ (estimated 
error radius of 50"), within the
WFC error circle.
This source was not detected in the ROSAT all sky survey (Voges,
private communication).
The image of the
field is shown in Fig.1 along with the refined WFC
error region (radius $\sim 1.9'$, \cite{Z97}). 
The previously known ROSAT source 1RXSJ0653.8+7916, lying
outside the WFC error box, was also detected.
 

\begin{figure*}
\centerline{
\includegraphics{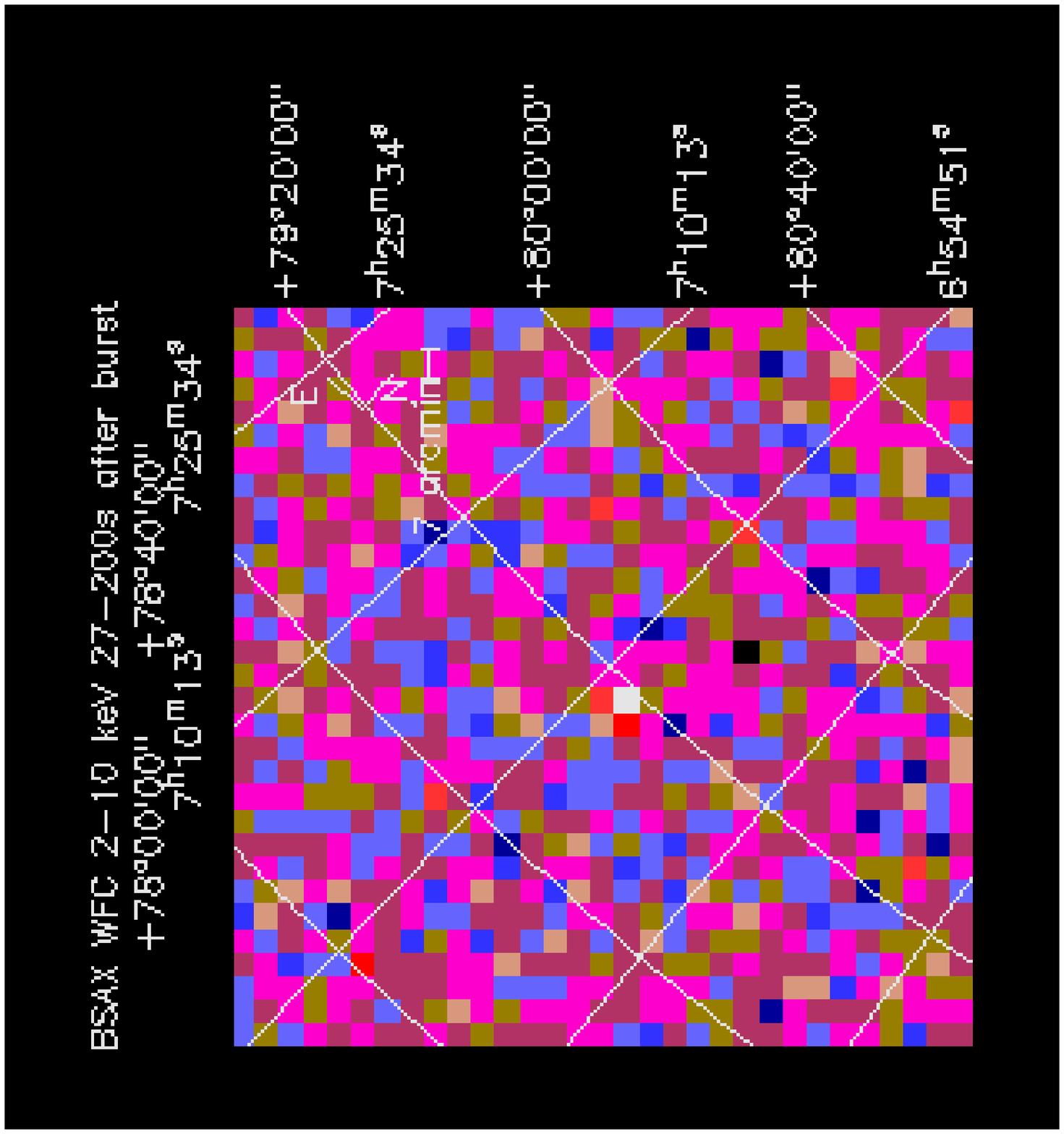}
\includegraphics{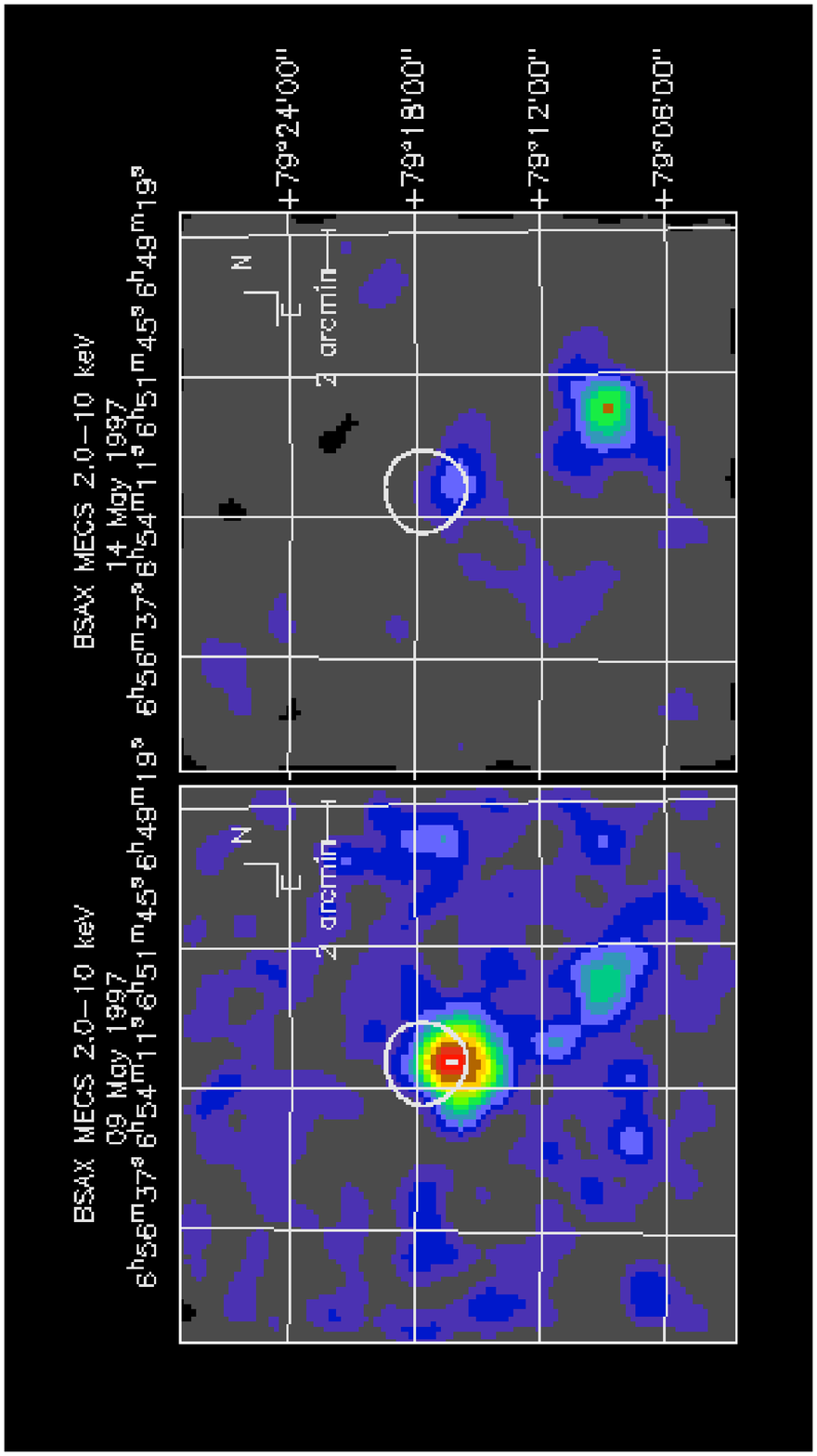}
}
\vskip 7truecm
\caption[]{Time sequence of images of the field of GB970508 observed by the 
WFC2 (left image, 27$-$200 s after the burst), MECS(2+3)
on May 9 (TOO1, center image, 6 hours after the GRB),  and 
MECS(2+3) on May
14 (TOO4, after 6 days). 
The WFC2 show the presence of the afterglow that was then
detected by the LECS and MECS 
(1SAXJ0653.8+7916 visible in the 99\% error circle of the WFC).
Note the decrease in intensity 
between the two MECS observations, as compared
to 1RXSJ0653.8+7916,  the source in the lower right corner}
\end{figure*}

Three other BeppoSAX observations (hereafter TOO2$-$4) 
were performed, the
last took place  $\sim 6$ days after the burst. 
 In all three observations we detected the source 
1SAXJ0653.8+7916  at a position consistent 
with that of the first observation.



\section{The association of 1SAXJ0653.8+7916 with GB970508}

The probability of finding a 
serendipitous X$-$ray source in the WFC error box 
with a flux greater than that observed is $\sim  10^{-3}$
(e.g Cagnoni et al. 1997).
 However, this probability should be revised  to consider the
potential association of 1SAXJ0653.8+7916
to classes of sources which show similar properties.
The very high value of X$-$ray emission compared to the optical
($\alpha_{ox}\sim 0.6$, where $\alpha_{ox}$ is the slope
of the power law $F\sim E^{-\alpha_{ox}}$ connecting the
optical to the 2 keV fluxes) is observed 
only in BL Lacs and emission line AGN (Maccacaro et al. 1980).
The latter association is 
excluded by the absence of strong emission lines in the  optical spectrum 
typical of AGN (\cite{MTZ97}).
The observed values of  
$\alpha_{ox}\sim 0.6$ and $\alpha_{ro}\sim 0.3$ (radio 
from Frail et al. 1997) are  the extreme of the range observed in X$-$ray selected BL Lacs (e.g. \cite{bllacs}), so even this potential association is
rather unlikely. Furthermore,
on the basis of the logN$-$logS (\cite{bllacx}),
we find that  the probability of a chance occurrence is $\sim 10^{-
3}$(see also \cite{CT97a}). This number takes into account the number
of searches in GRB error boxes - with an area comparable or less than
that of the WFC - by BeppoSAX (5) and other X$-$ray satellites (9).
Further support for  the association of the transient with the GRB 
afterglow comes from
the temporal behaviour observed in X$-$rays (next section) and 
other wavelenghts (e.g Djorgovski et al. 1997, Frail et al. 1997).

\section{Time evolution from the GRB to the afterglow}

GB970508 is a rather weak event, with a peak flux   
 $f_\gamma\sim 3.4 \times 10^{-7} \flux$ in the GRBM (40$-$700 keV) and
$f(2-26\ \rkeV)=(5.9\pm0.6)  \times 10^{-8}\flux$ in the WFC. 
In Fig. 2 we show the GRBM 
and WFC
light curves of the event. The event lasted about 15 s  in the GRBM and about
25 s in the WFC.  
The total fluence of the burst was $(1.8\pm0.3) \times 10^{-6} \rerg\ \rcm2$ and $(0.7\pm0.1) \times 10^{-6} \rerg\ \rcm2$ in the 
GRBM and WFC, respectively: 
about $40\%$ of the
burst energy emitted in the X$-$ray band. 
This fraction is substantially higher than that observed
in other bursts:
the value of
$\fxg\sim0.17$ of the peak fluxes is $\sim 5$ times greater than that 
of GB970228 (Frontera et al. 1997),
GB960720 (\cite{gb960720}) and the average value of the
GINGA sample (\cite{ginga,ginga2}). 

Costa et al. (1997a) attributed the train of pulses observed in GB970228
40 s after the initial burst to the beginning of the afterglow
(see also Frontera \etal 1997).
The light curves of GB970508 show
a second prominent pulse  
in  X$-$rays  $\sim 10$ s after the first one.
This is substantially softer than the first pulse but it 
merges with the first pulse, so we cannot 
conclude whether it represents the beginning of the afterglow
or not.
However, at $t>27$ s, faint residual activity  emerges from 
a detailed analysis of the WFC image (Fig.1). 
An analysis of the overall X$-$ray temporal behaviour 
supports the idea
that this
emission corresponds to the afterglow (Fig.3).
The decreasing flux observed by the WFC in the 27$-$200 s period
after the burst
connects to the first data points of TOO1 with
a power law $t^{-\delta}$ (where t is the time from the beginning
of the GRB)
 with $\delta=1.1\pm 0.1$.  

However, at $\sim 6\times10^4$ s, the
 flux increases  in
an outburst - on a time scale of
$\sim 10^5$ s - with a time behaviour similar
to that observed in the optical, followed by a 
sudden decrease observed in TOO4.
The latter is caused by a spectral steepening,
 that we will describe in more detail
in Piro et al. (1997d)

\begin{figure}
\epsfxsize=\hsize   \centerline{\epsffile{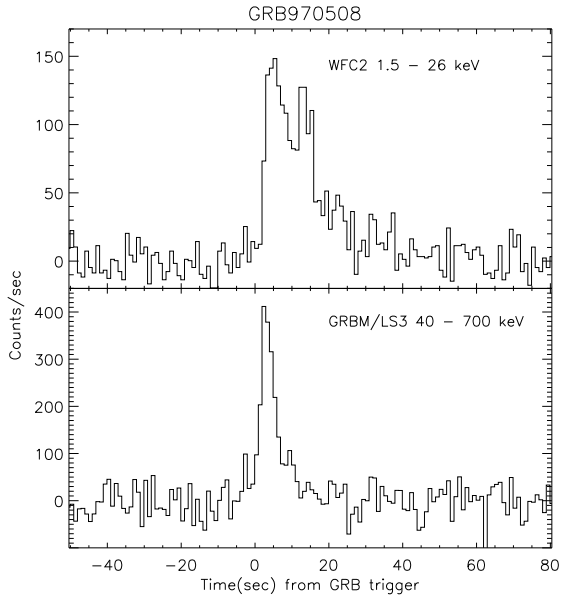}}
\caption[]{Light curves of GB970508 in the GRBM (bottom) and WFC
(top)}
\end{figure}

\begin{figure}
\epsfxsize=\hsize   \epsffile{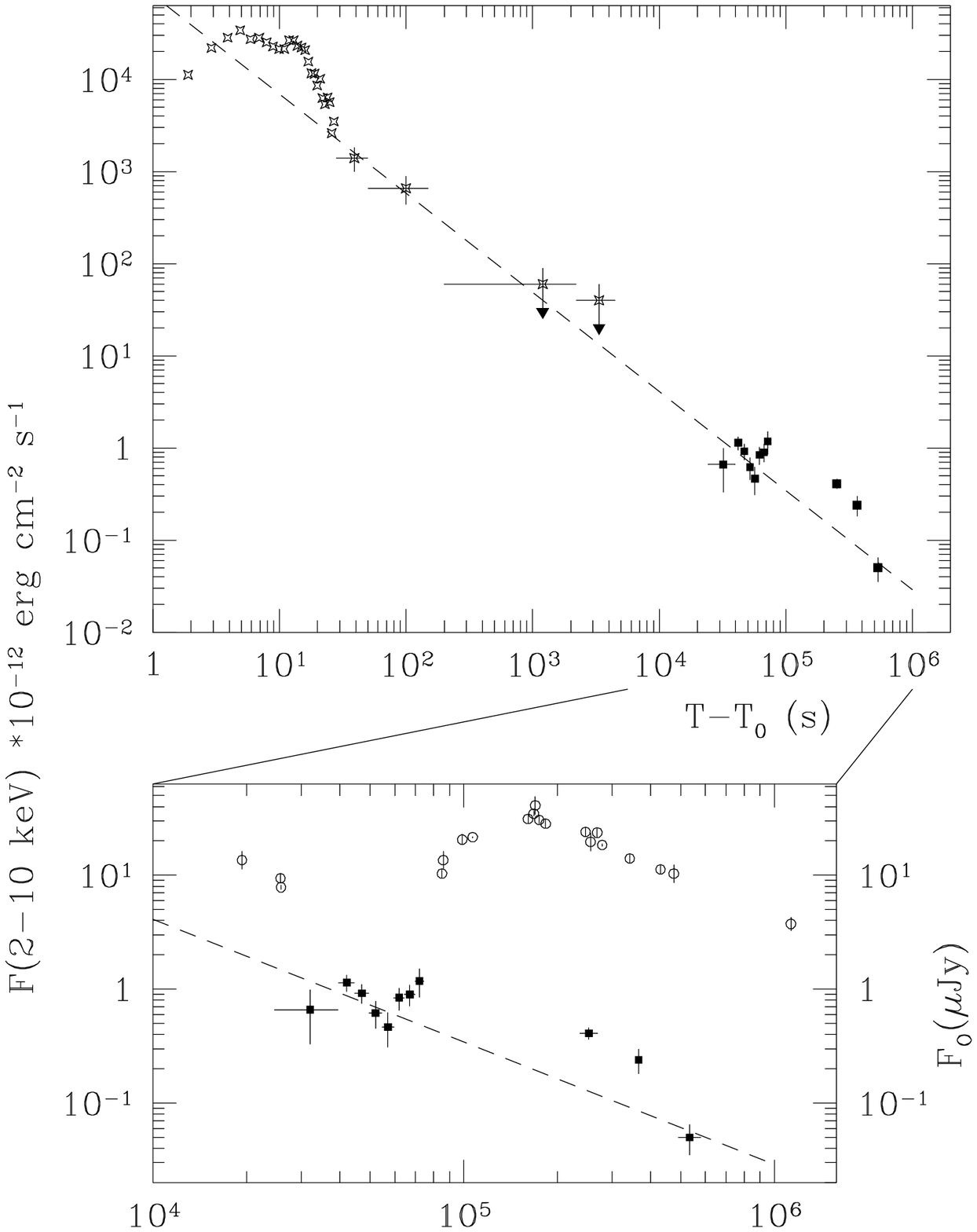}
\caption[]{Upper panel: X$-$ray light curve (2$-$10 keV) from the GRB to
the afterglow (upper panel, starred=WFC, filled squares=NFI).
The
dashed line is the best fit power law to  the WFC data (excluding the
GRB) and the first part of TOO1 data stream, before the increase
at $6\times 10^4$ s. The lower panel is a blown-up that includes
the optical behaviour of the source in the R band (open circles) from Galama et al. (1997), Castro-Tirado et al. (1997b), Chevalier \& Ilovaisky (1997), Mignoli et al. (1997), Schaefer et al. (1997), Groot et al. (1997), Garcia et al. (1997), Kopylov et 
al. (1997a,b). The vertical right hand scale refers to the optical data.
}
\end{figure}

\section{Discussion and Conclusions }

The combination of the WFC sensitivity and fast follow-up
with the NFI allowed to follow the evolution of the X$-$ray
emission of the GRB from 1 to $10^6$ s.  
We find that after the initial burst the X$-$ray 
emission detected with the WFC between 27 and 200 s, the two
WFC upperllimits between 200 and $4\times 10^{3}$ s and
the NFI measurements between $2.5\times 10^{4}$ and $6\times 10^{4}$ s 
are well fitted by a $t^{-1.1}$ power law decay. 
The data are therefore  consistent with a continuing
afterglow emission, although a deviation from the power-law decay, during the $200-4\times 10^{3}$ s time interval cannot be excluded. 
While this temporal behaviour is similar to that observed in GB970228
(\cite{c97nat}), GB970402 (\cite{gb970402}) and
GB970828 (\cite{gb970828}), it is the first time
that the afterglow was
detected immediately after the primary event. 
However, the evolution after $6\times 10^4 s$ deviates from this power law,
being dominated
by an outburst with a duration
$\sim  10^5$ s.
The energy released in the 2$-$10 keV range in the power law
component of the afterglow 
integrated from 27s  to $5.8\times 10^5$ s 
corresponds to about 20\% of the
total X- and $\gamma$-ray energy of the GRB. 
This is comparable to the case of gb970228 (\cite{c97nat}).
The energy excess with respect to the power
law during the burst event (from $6\times10^4$ s to
$5.8\times 10^5$ s), is $\sim 5\%$ of that of the
GRB.

Therefore, not only the afterglow 
carries an energy comparable to that of the main
event, but a significant fraction of this energy
is released in an outburst 
taking place on a time scale $\sim 10^4$ times larger 
than that of the GRB. 
The overall evolution of the afterglow {\it and} the GRB 
could be then described by a 
power law on the top of which bursts 
of different time scales occur, in particular on $1-10$ s (the
GRB proper) and on $\sim 10^5$ s.  
These results suggests that the same process is responsible for
both the GRB {\it and} the  afterglow. In the fireball
shock scenario (e.g \cite{RM97}, \cite{V97}, \cite{KP97}), models
in which both the GRB and the afterglow are produced by the
same mechanism 
 are therefore preferred.
The increase of the bursting duration with time agrees with the
general fireball scenario, where the timescales are primarily
determined by the superluminal motion of a shell,
whose Lorentz factor decreases very rapidly as the shell expands.

The optical turn up (Fig.3, lower panel) appears to
follow the X$-$ray burst with no substantial delay
(lag $< 2\times 10^4$ s), suggesting a same origin
for the optical and X$-$ray events. It then appears unlikely
that the optical turn up  is  produced by an energy dependent
effect, as a shift of the break energy (\cite{V97}, \cite{KP97}).

The reason of the different evolution of GB970508 compared to
GB970228  after the initial  phase is not clear. 
It could be
associated with the very soft primary event of GB970508
or with a different environment in which the fireball expands. This may also be the case of GB970111, for which there are indications of very faint afterglow activity (Feroci et al. 1997). 
It is however possible that similar bursts happened in
the other GRB's but have
been missed due to the sparse sampling of the light curves.

\begin{acknowledgements}
These results have been obtained thanks to
the extraordinary efforts of of the BeppoSAX team.
We would like to thank in particular L. Scarsi and G.C. Perola for
help in deciding the strategy of follow-up observations, 
the Mission Planners D. Ricci, M. Capalbi, S. Rebecchi,
the SOC scientists A. Coletta, G. Gandolfi, M. Smith, A. Tesseri, V, Torroni
, G. Spoliti,
the operation team L. Salotti, C. DeLibero and G. Gennaro for the fast re-pointing,
L. Bruca and G. Crisigiovanni for the fast processing of raw data, 
and P. Giommi, F. Fiore of SDC for helping in quick analysis of TOO data.
\end{acknowledgements}

\end{document}